\documentclass[review,number,sort&compress]{elsarticle}
\usepackage{lineno}
\linenumbers

\usepackage{graphicx}
\usepackage{amssymb}
\usepackage{color}
\usepackage{url}
\usepackage{ulem}
\usepackage[figuresright]{rotating}

\journal{Nuclear Instruments and Methods A}

\begin{document}

\begin{frontmatter}


\title{Monte Carlo Simulation of the Photon-Tagger Focal-Plane Electronics at the MAX IV Laboratory}


\author[illinois]{L.~S.~Myers}\fnref{fn1}
\author[gwu]{G.~Feldman}
\author[lund]{K.~G.~Fissum\corref{cor1}}
\ead{kevin.fissum@nuclear.lu.se}
\author[max4]{L.~Isaksson}
\author[kentucky]{M.~A.~Kovash}
\author[illinois]{A.~M.~Nathan}
\author[sal]{R.~E.~Pywell}
\author[max4]{B.~Schr\"{o}der}

\address[illinois]{University of Illinois at Urbana-Champaign, Urbana-Champaign IL 61802, USA}
\address[gwu]{The George Washington University, Washington DC 20052, USA}
\address[lund]{Lund University, SE-221 00 Lund, Sweden}
\address[kentucky]{University of Kentucky, Lexington KY 40506, USA}
\address[sal]{University of Saskatchewan, Saskatoon SK Canada S7N 5E2}
\address[max4]{MAX IV Laboratory, Lund University, SE-221 00 Lund, Sweden}

\cortext[cor1]{Corresponding author. Telephone:  +46 46 222 9677; Fax:  +46 46 222 4709}
\fntext[fn1]{present address: Duke University, Durham NC 27708, USA}

\begin{abstract}
Rate-dependent effects in the electronics used to instrument the tagger focal 
plane at the MAX IV Laboratory have been investigated using the novel approach 
of Monte Carlo simulation. Results are compared to analytical calculations as 
well as experimental data for both specialized testing and production running
to demonstrate a thorough understanding of the behavior of the detector system.
\end{abstract}


\end{frontmatter}

\section{Introduction}
\label{section:introduction}

The MAX IV Laboratory~\cite{m4} is the Swedish National Electron Accelerator 
Facility located in Lund, Sweden. The Tagged-Photon Facility (TPF) at the
MAX IV Laboratory~\cite{adler2012,tpf} is the beamline at the facility which 
utilizes the 200~MeV pulse-stretcher mode~\cite{lindgren2002}. Electrons in 
pulses with widths of about 200~ns are accelerated to energies up to 200~MeV 
and then injected into the MAX~I pulse-stretcher ring (PSR) at a frequency of 
10~Hz.  The electrons are then slowly extracted over the following 100~ms, 
before the arrival of the next pulse from the linac, and then transported to 
the TPF. In this manner, the pulsed electron beam originating in the injector 
is converted into a continuous, but non-uniform, electron beam
with an average current of approximately 20~nA.  
Understanding rate-dependent effects in the experiment
electronics resulting from the time structure in the electron beam required 
the efforts reported upon in this paper.

\begin{figure}
\resizebox{1.00\textwidth}{!}{\includegraphics{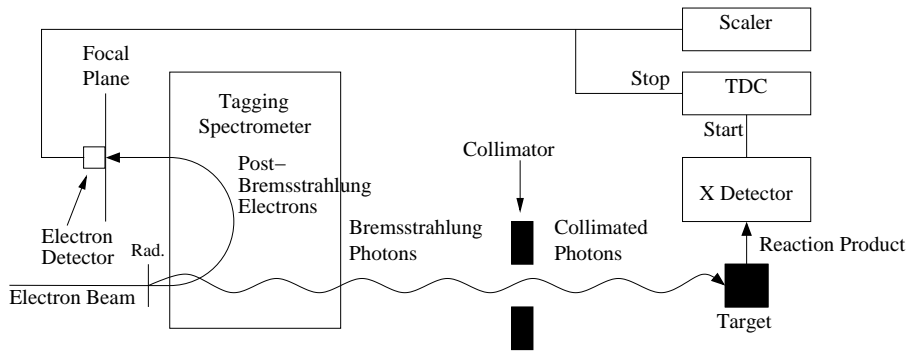}}
\caption{\label{figure:figure_01_tagging_technique}
The photon-tagging technique. Beam electrons may radiate bremsstrahlung
photons. Post-bremsstrahlung electrons are momentum analyzed using the
photon tagger, which consists of a magnetic tagging spectrometer and a FP array
of electron detectors. Bremsstrahlung photons which pass through the collimator
strike the experiment target and may induce photonuclear reactions which result
in a reaction product being detected. The coincidence between a reaction
product and a post-bremsstrahlung electron is a tagged-photon event.
}
\end{figure}

The TPF houses two photon-tagging spectrometers inherited from the Saskatchewan
Accelerator Laboratory (SAL) in Saskatoon, 
Canada~\cite{vogt1993,sal1994,sal1995}. 
These devices are used to perform photonuclear experiments via the well-known 
photon-tagging technique~\cite{adler1990} illustrated in  
Fig.~\ref{figure:figure_01_tagging_technique}. The 
electron beam passes through a thin metal radiator 
($\sim$100~$\mu$m Al), and a small portion ($\sim$0.1\%) of the incident 
electron beam is converted into a bremsstrahlung photon beam. Electrons that 
do not interact are dumped onto a well-shielded Faraday cup which registers the 
non-interacting electron-beam current. The resulting bremsstrahlung photon beam 
passes through a collimator to define its size prior to striking the 
experimental target. Post-bremsstrahlung electrons are momentum-analyzed using 
one of the magnetic photon-tagging spectrometers together with a 63-counter 
plastic-scintillator array positioned at the spectrometer focal 
plane (FP)~\cite{adler2012}. A time coincidence between a reaction product from 
a photon-target interaction and a recoiling electron is a tagged-photon event.

The energy of a tagged photon is determined from the difference between the 
energy of the incident electron beam and the energy of the post-bremsstrahlung 
electron detected in the focal plane.  The number of electrons striking a given
channel in the FP array is a crucial experimental parameter, as it is part of 
the overall experimental photon-flux normalization. The number of electrons 
must be corrected for the tagging efficiency~\cite{adler1997}, which 
measures the probability that the bremsstrahlung photon in question passes 
through the beam-defining collimator and is incident upon the experimental 
target.

The number of electrons counted in a given FP channel must also be 
corrected for rate-dependent effects. These effects arise due to the fact that 
the electron beam striking the radiator is not truly continuous, but rather has
a periodic structure of varying intensity. These intensity variations are due 
to the non-uniform filling of the PSR by the injector and the frequency of the 
shaker in the PSR which is used to disturb the 
electrons from the central orbit of the ring lattice in the extraction process.
The measure of this intensity variation is the duty factor of the 
beam~\cite{florizone1994}. At an operating current of 20~nA, the average rate 
in a FP channel is approximately 1~MHz; however, the instantaneous rate 
in the same FP channel can be as high as 4~MHz~\cite{myers2010} due to the duty
factor. Using this exact tagging spectrometer and FP detector array at such 
high rates, Hornidge~{\it et al.} observed substantial rate-dependent effects 
in deuterium Compton-scattering data measured 
at SAL~\cite{hornidge1999,hornidge2000}. In order to determine the necessary 
rate-dependent corrections, Pywell developed a Monte-Carlo simulation of the 
tagger setup~\cite{pywell2009}.  This original simulation has been completely 
overhauled and adapted to the present experimental conditions of the TPF at 
the MAX IV Laboratory.

In this paper, we present a detailed comparison between both dedicated test 
data and experimental production data obtained using the TPF and the overhauled
Monte Carlo, and we demonstrate a clear understanding of the rate-dependent 
effects that we have encountered.

\section{Rate-dependent effects} 
\label{section:rate_dependent_effects}

\begin{figure}
\resizebox{1.00\textwidth}{!}{\includegraphics{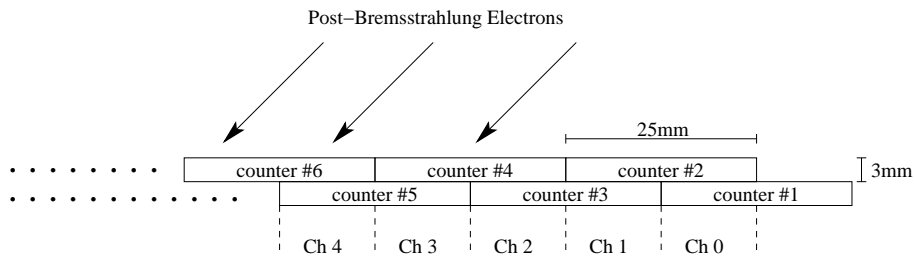}}
\caption{\label{figure:figure_02_focal_plane_hodoscope}
The FP hodoscope in 50\%-overlap configuration. A coincidence between
a counter in the front plane (even) and a counter in the back plane (odd)
defines a tagger channel. There are a total of 63 counters and 62 channels in the focal plane.
}
\end{figure}

The FP detector hodoscope consists of an array of NE110 scintillators arranged 
in two parallel rows of scintillators. The front row nearest the 
exit window of the tagger magnet has 31 elements, while the back row has 32 
elements. The relative orientation of the two rows may be varied in order to 
adjust the overlap between the rows from complete (100\%) to 50\% (the 
configuration used for this work). By decreasing the overlap, the 
recoil-electron energy resolution is increased and consequently the photon-energy resolution is 
also increased (see Fig.~\ref{figure:figure_02_focal_plane_hodoscope}).

\begin{figure}
\resizebox{1.00\textwidth}{!}{\includegraphics{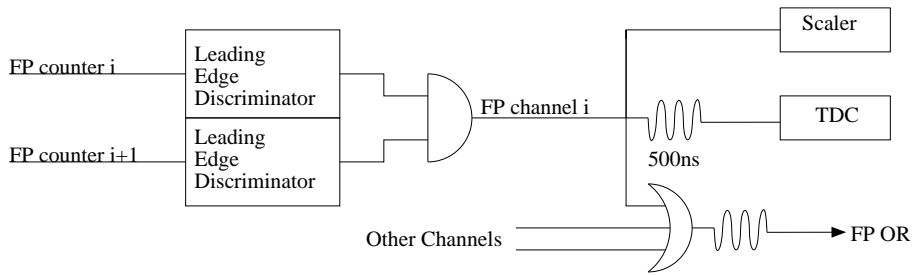}}
\caption{\label{figure:figure_03_focal_plane_electronics}
The FP electronics. A coincidence between an electron-detector signal in the
front plane and an electron-detector signal in the back plane defines a tagger
channel. The coincidence module looking for these overlaps was from SAL. This
signal is counted and used to stop a TDC started by the photonuclear
reaction-product detector. The logical OR of all the focal-plane channel
signals is the FP~OR trigger.
}
\end{figure}

Each counter is instrumented with a Hammamatsu R1450 photomultiplier tube with 
a 19~mm head, and high voltage is supplied by a LeCroy 1440 power supply.
The signals from the FP counters are passed to LRS~4413 leading-edge 
discriminators operated in burst-guard mode. These discriminators are used to 
generate logic signals of widths varying from 25~ns to 50~ns depending upon 
the experiment in question. Coincidences between two overlapping scintillators 
in the front and back rows are identified in overlap coincidence modules 
designed and built at SAL -- it is these coincidences that define tagger 
channels. When operated in 50\% overlap mode, the length of the hodoscope is 
842~mm, and each of the so-defined tagger channels has a physical width of 
13~mm. The logical OR of the 62 FP channels is used as a trigger for a 
recoil-electron event (see 
Fig.~\ref{figure:figure_03_focal_plane_electronics}). When a recoil-electron 
trigger occurs in coincidence with a trigger from the experiment detectors, a 
candidate tagged-photon event is registered. 

\begin{figure}
\resizebox{1.00\textwidth}{!}{\includegraphics{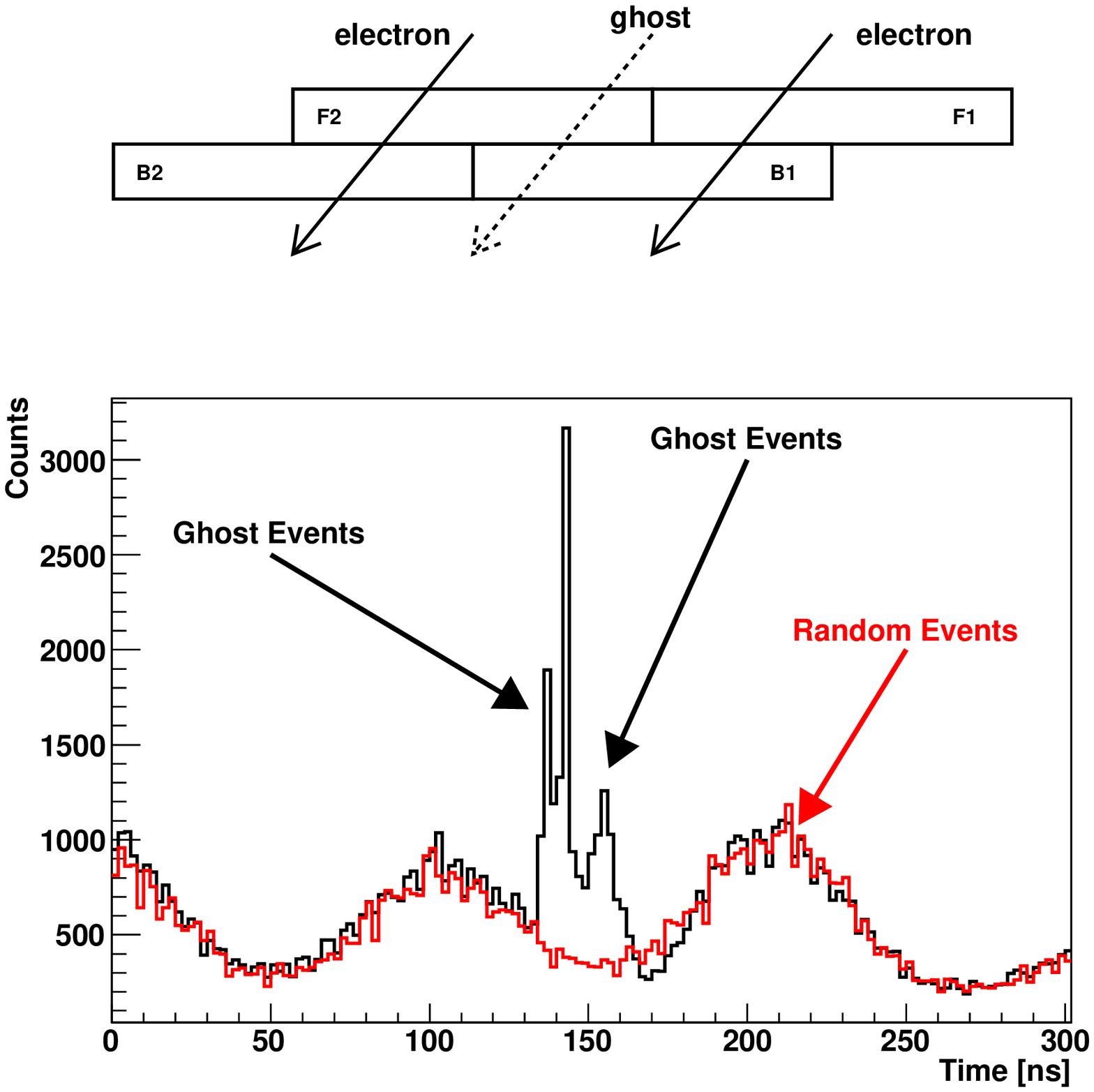}}
\caption{\label{figure:figure_04_ghosts}
(Color online) (Top) The scenario that leads to a ghost event. Real electrons 
(solid arrows) in next-neighboring FP channels arrive nearly simultaneously. 
This creates the illusion -- or ``ghost'' -- of an electron 
(dashed arrows) in the counters that constitute the intermediate FP channel.  
(Bottom) A demonstration of ghost events within the data.  A 
coincidence between the experiment trigger and FP channel 36 has been required.
Adjacent FP channel 37 (black) and physically distant FP channel 40 (red) TDC 
spectra are shown. Both were started by the same experiment trigger. The 
red spectrum is thus due to untagged, accidental events. The ``ripple" in
the accidental distribution is discussed in 
Sect.~\ref{subsection:electron_beam_profile}. The black spectrum also shows 
these untagged, accidental events. In addition, there are three clear peaks
between 130 and 170~ns. The sharp, dominant peak is a combination of 
accidental coincidences and ghosts.  The secondary and tertiary peaks to the
left and the right are due to ghost events. The multiple peaks were determined 
to arise from the pulse-timing characteristics of the TDC (see 
Sect. \ref{subsection:electronics}).
}
\end{figure}

Two advantages to requiring a coincidence between the front and back rows of 
scintillators in the FP hodoscope are that the gamma-ray background 
in the tagger hall is not registered by the tagger, and that the photon-energy 
resolution may be increased by varying the relative orientation of the two 
scintillator planes rather than building a new array with smaller 
scintillators. However, a major disadvantage of requiring a coincidence between
the front and back rows of scintillators in the FP hodoscope is the creation of
so-called ``ghost events'' at high rates. 
The ghosts are an artificial creation of the instrumentation of the focal 
plane.
The scenario leading to a ghost event is illustrated in the
top panel of Fig.~\ref{figure:figure_04_ghosts}. Two electrons strike 
next-to-neighboring channels (counters F1 $\cdot$ B1 and F2 $\cdot$ B2) 
simultaneously which creates the illusion of an electron in the channel 
in the middle (counters F1 $\cdot$ B2) -- the ghost event.
The accidental coincidences that result in ghost events depend on 
the electron rate, the width of the the discriminator output pulses, 
and the resolving time of the coincidence modules. Because these ghosts are 
formed in the FP electronics, the accidental coincidences are registered in 
both the scalers and the TDC modules. This results in a partial cancellation of
the effect. Understanding the dependence of the ghosts on the input conditions 
and determining the amount of cancellation requires a Monte Carlo simulation of
the tagger setup and electronics.

\begin{figure}
\resizebox{1.00\textwidth}{!}{\includegraphics{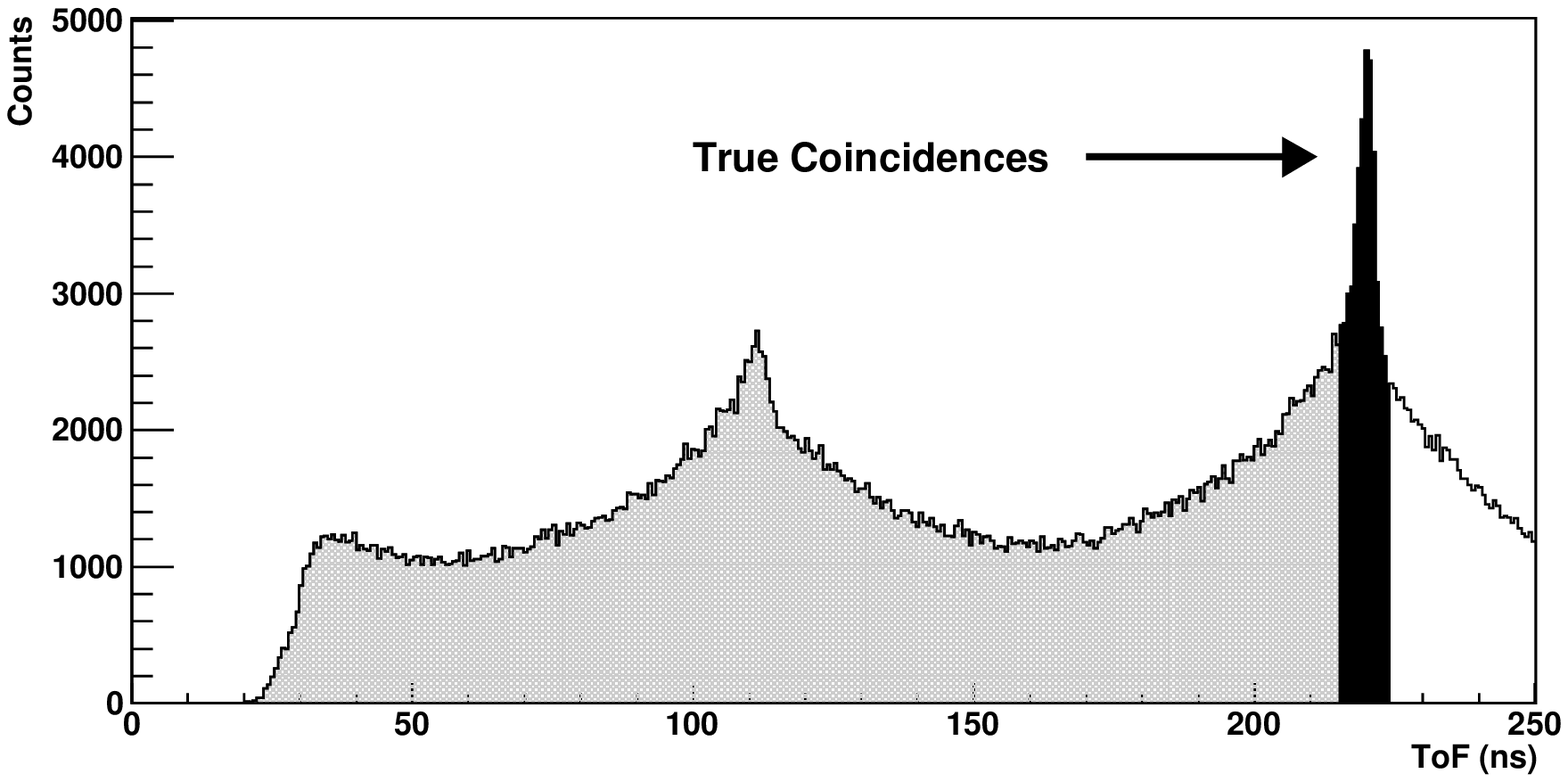}}
\caption{\label{figure:figure_05_stolen_coincidences}
An illustration of the stolen-coincidence effect in a single-hit TDC spectrum
acquired at a high post-bremsstrahlung electron rate. The black peak at channel
225 represents true coincidences between the reaction-product detector and the
hodoscope. This point in time is the earliest possible that a true
coincidence may be registered. Events in the grey shaded region correspond to
random post-bremsstrahlung electrons stopping the FP TDCs before this
earliest possible point in time. The true coincidence is thus stolen when a
single-hit TDC is used. Note that the ``peak'' at channel 110 is due to the 
non-uniform filling of the ring in the extracted electron beam.
}
\end{figure}

High recoil-electron rates can also lead to real electron stops being missed 
by the leading-edge discriminators (even though they are operated in 
burst-guard mode) and the overlap coincidence modules due to dead-time effects. 
They also result in asymmetries in the data acquisition, since the scalers 
counting the electron signals which are used to normalize the data are much 
faster than the TDCs which identify tagged-photon events -- see Sect.
\ref{subsection:electronics}. And finally, when single-hit TDCs are employed, 
a random electron may be detected in the FP channel before the actual
electron that corresponds to the tagged photon. The result is that the 
single-hit TDC stops too early, leading to a well-studied phenomenon known as 
stolen coincidences~\cite{owens1990} -- see 
Fig.~\ref{figure:figure_05_stolen_coincidences}.

\begin{figure}
\resizebox{1.00\textwidth}{!}{\includegraphics{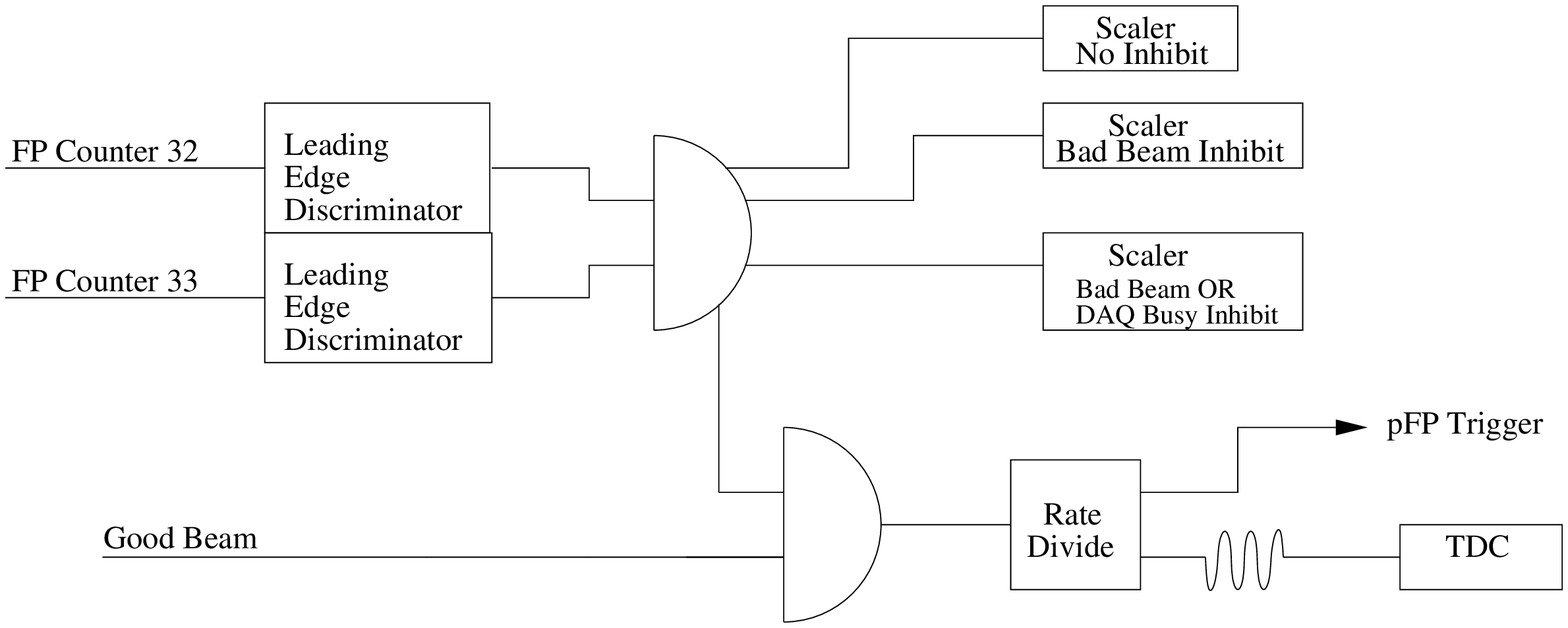}}
\caption{\label{figure:figure_06_pFP}
The pFP trigger. The signal from a single FP channel is rate divided to
$\sim$10~Hz and then passed to the DAQ as a valid event trigger and used to
start the FP TDCs.
}
\end{figure}

In order to measure the number of stolen coincidences in the data set, a 
prescaled FP (pFP) trigger is routinely included in the data stream --
see Fig.~\ref{figure:figure_06_pFP}. To form this trigger, a representative 
logic signal from a single FP channel is passed to a bank of three scalers 
(uninhibited, bad-beam inhibited\footnote{Bad beam is defined as the first 
1~ms of the extracted beam from the PSR.}, and bad-beam OR busy DAQ inhibited).
A fourth copy of this signal is bad-beam inhibited, rate-divided (typically 
by a factor of 10$^5$), and used as the pFP trigger. It is trickled 
into the data stream at a rate of $\sim$10~Hz as a valid trigger and is 
used to start the FP TDCs. As a result, for the pFP-triggered subset of 
events, the FP TDCs are started by the pFP signal and stopped by the
usual recoil-electron signal (recall 
Fig.~\ref{figure:figure_03_focal_plane_electronics}). Note that the coincidence 
between the normal recoil-electron signals in the front and back scintillator 
planes was established using SAL modules, while the same coincidence for the 
pFP trigger was established using a LRS~622.

\begin{figure}
\resizebox{1.00\textwidth}{!}{\includegraphics{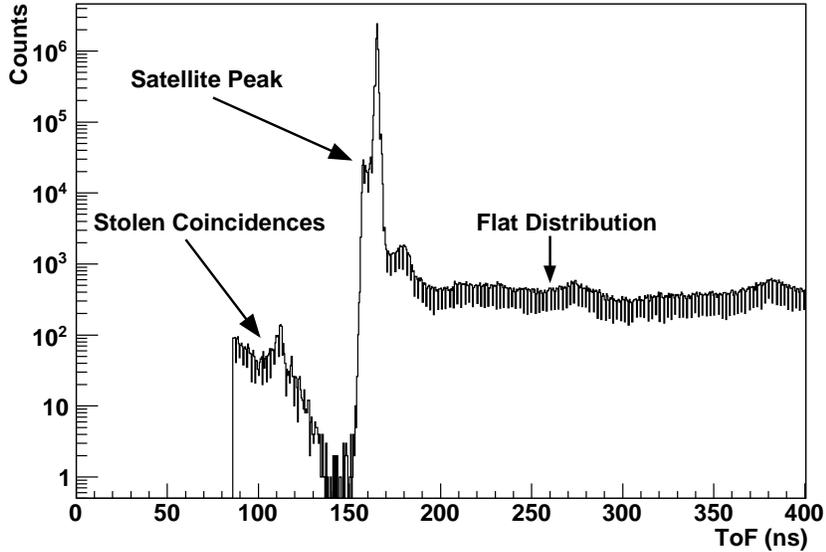}}
\caption{\label{figure:figure_07_pFP}
A typical pFP TDC spectrum. A self-timing peak between channels 165 and 175 is
clearly evident. The grouping of events below channel 150 are stolen
coincidences -- see text for details. The satellite peak between channels 150
and 165 as well as the flat distribution of events from channels 180 to 400
and higher could not be explained without the help of the simulation.
}
\end{figure}

Figuse~\ref{figure:figure_07_pFP} shows a typical TDC spectrum where the 
start signals were provided by the pFP trigger and the stop signals were 
provided by recoil electrons striking the focal plane. In the TDC spectrum, 
we expected to see a single self-timing peak 
(shown here from channels 165 to 175) and a grouping of events to the left of 
the self-timing peak (shown here below channel 150) corresponding to stolen 
coincidences. The appearances of a satellite peak (shown here from channels 150 
to 165) as well as a flat distribution of events at times greater than the 
self-timing peak (events ranging from channel 180 to 400 and higher) were a 
mystery prior to the simulation efforts.

\section{Basic features of the simulation}
\label{section:basic_features_of_the_simulation}

Using the procedure established by Pywell, the Monte Carlo simulation models 
the FP electronics in 1~ns steps from the recoil-electron detectors to the 
scalers and TDCs, traversing all of the intermediate electronics in the 
process.  In contrast to the Pywell procedure, our simulation addresses all 62 
FP channels simultaneously. In an effort to make the simulation as versatile as 
possible, it was written so that many of the initial conditions are input 
from separate files rather than being hard-coded. During each ns of the 
simulation, the FP channels are checked to see if a recoil electron struck 
the channel. The probability for such an electron event is dependent on the 
instantaneous electron rate in the FP channel, which in turn depends on the 
time structure of the electron beam. Both of these parameters are adjustable 
inputs. 

A Poisson distribution is used to generate recoil electrons in the FP 
channels\footnote{
The individual FP counter rates were not recorded.
}. If an electron is observed in a given channel, the corresponding 
counter discriminators are updated, and then the counter discriminator pulses 
are tracked to the overlap coincidence modules which are updated. The signals 
from the overlap coincidence modules are then propagated 
to the FP scalers and TDCs which record the electron. For each recoil electron 
in a FP channel, there is also a probability that the corresponding photon 
generates a start for the FP TDCs. The likelihood of this occurring (which
is related to the tagging efficiency) is another input to the simulation. 
Additionally, untagged photons and cosmic-ray events can also be used to start 
the FP TDCs\footnote{
The corresponding count rate is used to generate a random event using a 
Poisson distribution.}. 
If there is an experiment trigger (either a pFP electron, a tagged or untagged 
photon, or a cosmic ray), the FP TDCs are started and the FP scalers 
are inhibited. The duration of the inhibit may be adjusted to precisely match 
the experimental running conditions. Finally, the simulation checks each 
coincidence in the coincidence modules to see if a corresponding real electron 
is responsible for generating the coincidence. If no electron was present, 
then a flag is set which identifies the coincidence output as a ghost event. 
This flag allows for the analysis of both real and ghost events and leads to 
the subsequent correction factors for the data.

\begin{figure}
\resizebox{1.00\textwidth}{!}{\includegraphics{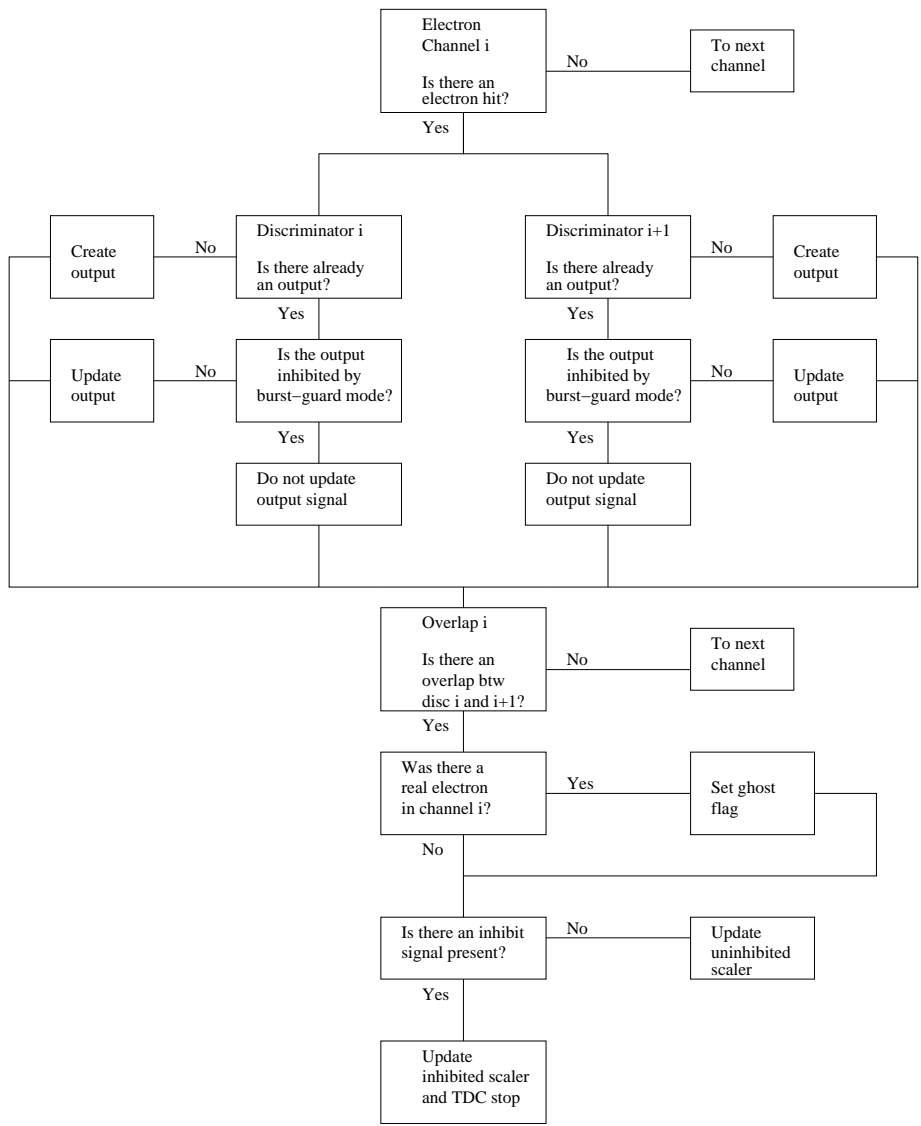}}
\caption{\label{figure:figure_08_flowchart}
Flowchart describing the simulation. See text for details.
}
\end{figure}

The original code developed at SAL by Pywell was written in \textsc{fortran}. 
The new simulation was updated to \textsc{C++} and was compiled with 
\texttt{gcc}, and the simulation output is written to a \textsc{root} 
file~\cite{root}. The compiled code is run on a 2.66~GHz CPU which takes 
approximately one hour to simulate the FP electronics for events spanning a 
time interval of one second. A flow chart describing the simulation is shown 
in Fig.~\ref{figure:figure_08_flowchart}, and a summary of input parameters 
is presented in Table~\ref{table:table_01_simulation_input_parameters}. Most 
of the inputs to the simulation listed in 
Table~\ref{table:table_01_simulation_input_parameters} are taken directly 
from the electronics setup (such as pulse widths) or from the data itself 
(such as electron rates and tagging efficiency). 

\section{Input parameters}
\label{section:input_parameters}

\subsection{Electronics}
\label{subsection:electronics}

Once the framework of the tagger simulation was developed, it was necessary to 
understand the response of the various electronics modules in order to 
implement them correctly. This required a thorough understanding of the 
leading-edge discriminators, various coincidence modules, TDCs, and scalers.

As previously mentioned, the LRS~4413 leading-edge discriminators used to 
instrument the focal plane were operated in 
burst-guard mode, which resulted in behavior that was 
especially important at high rates. The modules are designed to produce an 
output pulse of a fixed, user-determined width when the threshold is crossed. 
If a second event arrives at the discriminator before the end of the
previous output pulse, the output will be updated and will extend to the 
greater of the fixed width or the threshold re-crossing of the second input 
pulse. Any subsequent inputs will be ignored until the fixed width has passed 
and the discriminator is reset.

\begin{figure}
\resizebox{1.00\textwidth}{!}{\includegraphics{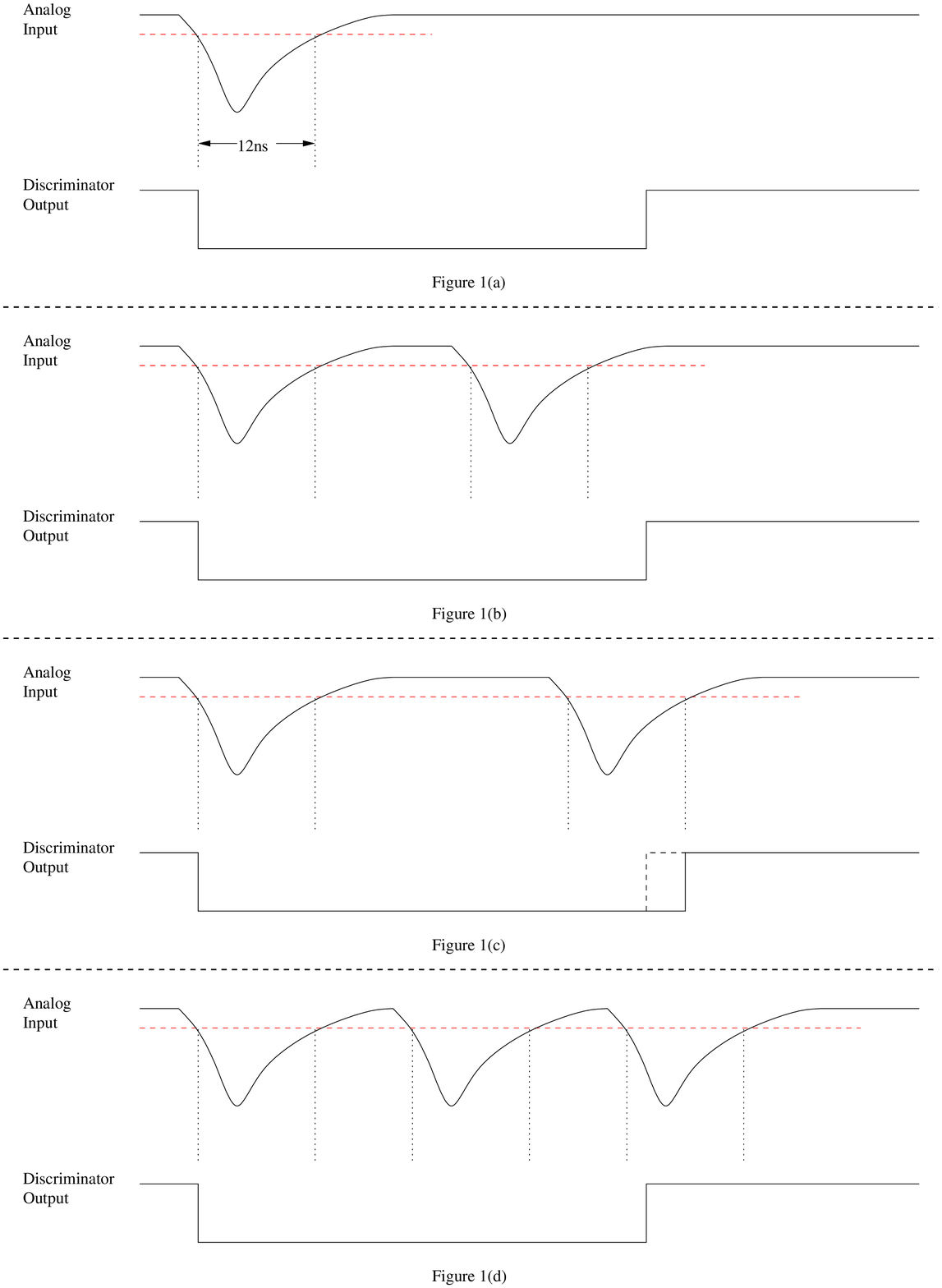}}
\caption{\label{figure:figure_09_burstguard}
An illustration of the behavior of the LRS~4413 leading-edge discriminators 
operating in burst-guard mode. See text for details.
}
\end{figure}

Figure~\ref{figure:figure_09_burstguard} illustrates the burst-guard behavior 
of the LRS~4413 leading-edge discriminator. In panel (a), a single analog 
signal corresponding to an electron is discriminated, resulting in a 40~ns 
output pulse. Panels (b) and (c) show the conditions for updating the output 
pulse with a second electron. In panel (b), the re-crossing associated with the
second electron occurs before the original output ends -- the output pulse is 
thus the same as in panel (a), and the second electron is missed. In panel (c),
the second re-crossing occurs after the output would have ended, so the 
duration of the output signal is extended. Third hits are never registered, 
as shown in panel (d).

The SAL overlap coincidence modules used for forming the pulses that were sent 
to the FP TDCs and scalers were continuously updating. An output pulse was 
generated whenever the two input pulses overlapped and was terminated whenever 
one or both inputs were reset. An overlap of at least 3~ns was 
necessary to produce an output pulse. This behavior is asymmetric to the 
LRS~622 coincidence module used to generate the pFP
trigger -- it produced a 
fixed output pulse that began when the discriminator output signals from the
two FP counters overlapped.
The CAEN~V775 single-hit TDCs used to instrument the focal
plane were experimentally determined to require input pulses of at least 
11~ns in width in order to register the pulses. In contrast, 
the CAEN 830 scalers which counted the recoiling electrons 
were experimentally determined to require input pulses of 
at least 3~ns in width in order to register the pulses\footnote{
Note that this $\sim$3~ns width is comparable to the smallest pulse that can be 
generated by the coincidence modules, so that the scalers may register even 
shorter pulses.
}. 

\begin{figure}
\resizebox{1.00\textwidth}{!}{\includegraphics{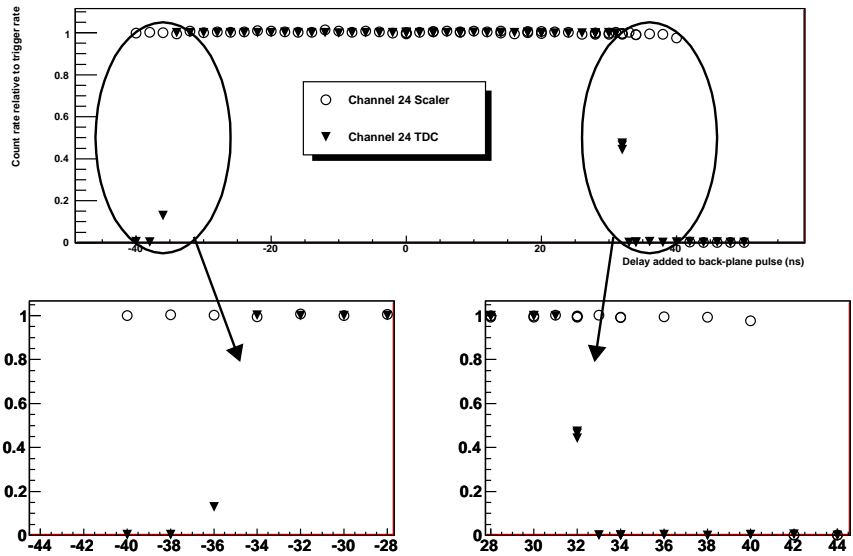}}
\caption{\label{figure:figure_10_overlap_test_result}
A comparison of the response of FP channel 24 TDCs and scalers when 
the pulses from the front-scintillator plane and the back-scintillator plane 
are set to 45~ns in width and the relative timing of the back-plane pulses is 
then varied by $\pm$40~ns.  Open circles correspond to the scaler and filled 
triangles correspond to the TDC. The bottom left panel is an enlargement of 
the top panel from $-$44~ns to $-$28~ns and the bottom right panel is an 
elnargement of the top panel from 28~ns to 44~ns. There is a clear asymmetry 
in the response of the two modules for shifts in relative timings of more than 
30~ns.
}
\end{figure}

Figure \ref{figure:figure_10_overlap_test_result} shows the count rate relative 
to the trigger rate for FP channel 24 for a wide range of relative 
pulse timings generated by a pulser. In this test, pulser triggers running at 
10~Hz were fed into the SAL overlap coincidence unit 
inputs for the front and back scintillator planes corresponding 
to FP channel 24. Tests were performed for  
pulse widths of 25~ns, 35~ns, and 45~ns. The relative timing between the 
front and back pulses was then varied in steps of a 
few ns over a total range of $\pm$40~ns. The open circles correspond to the 
FP channel 24 scaler, while the filled triangles correspond to the FP channel 
24 TDC.  Clearly, over the entire dynamic range investigated, the CAEN~V830 
scalers were able to register the signals from the SAL overlap coincidence 
units. This was true even when the timing of the back-plane pulse relative to 
the front-plane pulse was artificially fixed 40~ns early or 40~ns late, 
resulting in an overlap pulse of $\sim$3~ns width. On the other hand, the 
CAEN~V775 single-hit TDCs only registered pulses when the timing of the 
back-plane pulse relative to the front-plane pulse was between 35~ns early and 
30~ns late, resulting in an overlap pulse of $\sim$11~ns width. 

The fact that a $\sim$3~ns overlap pulse could generate a pFP trigger and, 
consequently, a start signal for the FP TDCs but an $\sim$11~ns overlap pulse 
was needed to generate a stop signal resulted in missing stops. The flat 
distribution from channels
180 to 400 in Fig.~\ref{figure:figure_07_pFP} thus resulted from random 
electrons stopping the FP TDC.  Obviously, this important asymmetry in the 
response of the electronics must be taken into account and corrected in order 
to properly normalize experimental data. 

\begin{figure}
\resizebox{1.00\textwidth}{!}{\includegraphics{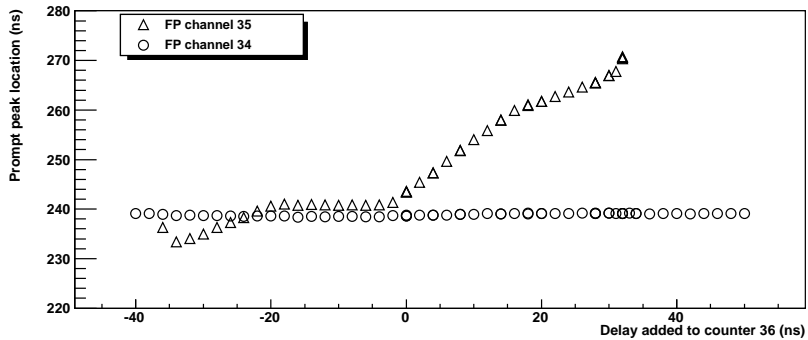}}
\caption{\label{figure:figure_11_ghost_timing}
Ghost timing. 
Recall Fig. \ref{figure:figure_04_ghosts}. The location of the 
overlap coincidence peak in the FP TDCs for channels 34 and 35 is 
shown as a function of varying the delay in FP counter 36. See text 
for details.
}
\end{figure}

A peculiar feature of the FP TDCs and/or overlap coincidence modules was 
observed during the test of the FP TDC resolving time. This test was 
performed using a 10~kHz pulser, and the signal from the pulser was split into 
three identical copies, which were then passed to the electronics connected to 
FP counters 34 (front plane), 35 (back plane), and 36 (front plane). 
Holding the timing fixed for the counter 34 and 35 electronics, 
it was observed that as delay was added to or removed from 
counter 36, the location of the counter 35/36 
(channel 34) overlap coincidence peak in the FP TDC varied by 
as much 
as 40~ns. A shift was expected to some degree (for example, for fixed counters 
34 and 35, as counter 36 is mistimed to arrive later and 
later, then the coincidence peak representing counter 35/36 overlaps should
also shift in time by the amount of the mistiming); however, the 
behavior shown in Fig.~\ref{figure:figure_11_ghost_timing} is not 
consistent with this behavior. 
There is a shift in the prompt
peak location at around $-$30~ns of added delay as well as non-linear   
behavior (a ``kink" and then a ``dogleg") above $+$20 ns. This variation in 
the location of the 
coincidence peak as a function of the relative timing of the front-plane and 
back-plane signals explains the ghost-peak structure in 
Fig.~\ref{figure:figure_04_ghosts} (recall the three peaks) and the satellite 
peak between channels 150 and 165 in Fig.~\ref{figure:figure_07_pFP}.
This effect was included in the simulation.

\subsection{Electron-beam profile}
\label{subsection:electron_beam_profile}

\begin{figure}
\resizebox{1.00\textwidth}{!}{\includegraphics{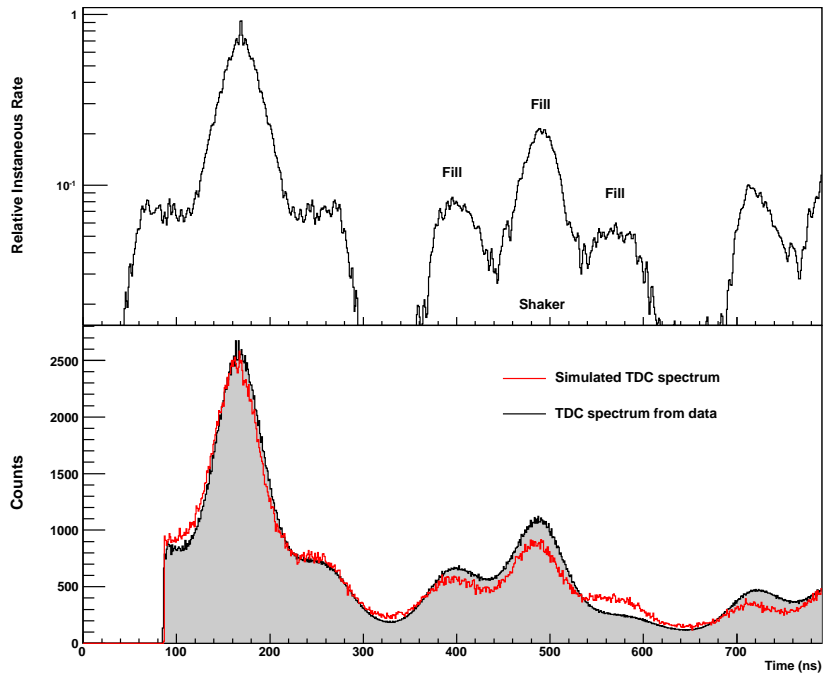}}
\caption{\label{figure:figure_12_beam_profile}
(Color online) (Top panel) The electron-beam time profile, $P(t)$. 
The observed time structure is due to incomplete filling of the 32.4~m 
diameter (108~ns) MAX~I PSR and the 3.3~MHz frequency (305~ns 
period) of the extraction shaker. The dropoff between the 
intensity of the first pulse and the second pulse is roughly a factor of 
four. This dropoff rapidily vanishes after the second pulse. The profile 
repeats throughout the 100 ms extraction of the beam. See text for details. 
(Bottom panel) A 
comparison between a simulated accidental TDC spectrum, $A(t)$, for the OR of all 62 
FP TDCs and data (shaded). The simulation input was the electron-beam 
profile shown in the top panel together with the parameters presented in 
Table~\ref{table:table_01_simulation_input_parameters}. Agreement is 
excellent, confirming our understanding of the input parameters.
}
\end{figure}

One very important input parameter to the simulation is the time profile of the
electron beam extracted from the PSR. Unfortunately, this is not directly accessible from the data. 
To obtain the time profile of the electron beam, the subset of the data 
pertaining only to pFP triggers was employed. This data set was generated by
requiring a recoil-electron signal in a specific FP channel. For this 
data set, recoil electrons striking FP channels well-separated from the 
selected channel were taken to be accidentals. The FP TDC distributions 
for these distant channels were then summed together to produce a purely 
accidental FP TDC spectrum, $A(t)$, with high statistics. 
This spectrum was 
then related to the true electron-beam profile via an auto-correlation 
function according to
\begin{equation}
\label{equation:auto_correlation_function}
A(t) = \int^{\infty}_{-\infty}P(T)P(t+T)dT.
\end{equation}
The method of Fourier transform was then used to extract the beam profile 
$P(t)$ shown in the top panel of 
Fig.~\ref{figure:figure_12_beam_profile}\footnote{
An array was created representing the accidental spectrum with one entry per 
ns. The FFT transform function in \textsc{root} was then used to produce 
$P^2(T)$. High-frequency terms deemed unimportant were filtered before taking 
the square root of $P^2(T)$ and performing the inverse FFT.}.
The secondary and tertiary time structures in this distribution result from the
3.3~MHz frequency (305~ns period) of the shaker used in the extraction of the 
electron beam and the incomplete filling of the 32.4~m diameter (108~ns period) MAX~I PSR.

\subsection{Instantaneous electron beam rates}
\label{subsection:instantaneous_electron_rates}

\begin{figure}
\resizebox{1.00\textwidth}{!}{\includegraphics{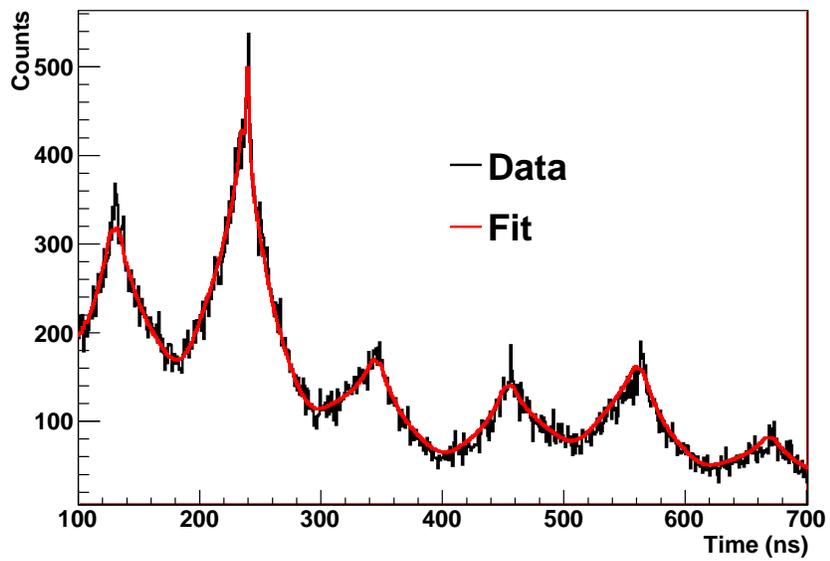}}
\caption{\label{figure:figure_13_fitted_TDC_spectrum}
(Color online) The result of fitting Eq.~\ref{equation:fitting_function} to
FP channel 17 TDC data (shown as a histogram). The time constant of the fit is 
employed in the simulation as the instantaneous electron rate.  The true 
coincidences occur at $t_0$ = 240 ns.
}
\end{figure}

The determination of the average instantaneous beam rate over one beam period 
(hereafter, the instantaneous rate) is crucial for an accurate calculation of 
both the stolen coincidences and the ghost corrections. The rate is given 
by the time constant of the accidental TDC spectrum which can be determined by
fitting the data with an exponential function. However, due to the complicated 
time structure of the beam, the method of extracting the rate must be carefully
chosen so as not to be sensitive to the fitting region selected. This is done 
by defining the fitting function as
\begin{equation}
\label{equation:fitting_function}
f(t) =
p_0~A(t)~e^{-Rt} + p_1~e^{(t-t_0)^2/2\sigma^2},
\end{equation}
where $p_0$ and $p_1$ are constants, $A(t)$ is the FP TDC spectrum for 
accidental events (see Eq.~\ref{equation:auto_correlation_function}), and $R$ 
is the instantaneous rate. Since the spectrum also contains some true 
coincidences, a Gaussian function centered at $t_0$ is used to account for 
these events so as to eliminate any bias they might cause in the fitting 
procedure. As seen in Fig.~\ref{figure:figure_13_fitted_TDC_spectrum}, this 
method accurately fits the data.

\small
\begin{sidewaystable}
\caption{\label{table:table_01_simulation_input_parameters}
Simulation input parameters. See text for details.}
\begin{center}
\begin{tabular}{ll} \hline \hline
Input parameter & Description \\
\hline
                     total run time              &                                     number of ns to run the simulation \\
              electron rate in reference channel &                                           taken from data, $\sim$1~MHz \\
                           cosmic-ray event rate &                                           taken from data, $\sim$50~Hz \\
                             untagged event rate &                                      taken from data, $\sim$50--150~Hz \\
                   probability of a tagged event &         taken from data, $\sim$50\% (inbeam), $\sim$0.5\% (scattering) \\
     electron rate relative to reference channel &                                            taken from data, $\sim$1--4 \\
           focal-plane discriminator pulse width &                                          leading edge, $\sim$25--50~ns \\
                                inhibit duration &                                       bad beam or DAQ busy, $\sim$1~ms \\
                                geometric double &                   recoil electron strikes adjacent channels, $\sim$1\% \\
                             pFP trigger channel &                                                             channel 32 \\
                         pFP trigger pulse width &         same as focal-plane discriminator pulse width, $\sim$25--50~ns \\
                          pFP overlap properties &                                             flag, module (non)updating \\
                                     pFP trigger &                                                           flag, on/off \\
                duty-factor meter (DFM) counters &                                              counter 20 and counter 50 \\
                                DFM pulse widths &                                                        $\sim$25--50~ns \\
                     DFM overlap characteristics &                                             flag, module (non)updating \\
                            prompt-peak location &          $\Delta t$ between FP TDC start and prompt peak, $\sim$100~ns \\
        periodicity of the electron-beam profile &                            long enough to generate several repetitions \\
                           electron-beam profile &              see top panel of Fig.~\ref{figure:figure_12_beam_profile} \\
\hline \hline
\end{tabular}
\end{center}
\end{sidewaystable}

\section{Comparison to data}
\label{section:comparison_to_data}

\subsection{Focal-plane TDC spectrum}
\label{subsection:focal_plane_tdc_spectrum}

Figure~\ref{figure:figure_12_beam_profile} (bottom panel) shows a comparison 
between the simulated FP TDC spectrum for accidentals for the OR of all 62 FP 
TDCs and data. The inputs included the electron-beam profile shown in the top 
panel of Fig.~\ref{figure:figure_12_beam_profile} together with all the 
parameters presented in Table~\ref{table:table_01_simulation_input_parameters}. 
A correction based on the location of the coincidence peak 
has been applied to align all of the individual 62 FP TDC spectra. 
The broad peaks in this spectrum are due to the structures in the 
electron beam extracted from the PSR. Time periods when greater numbers of 
electrons are extracted from the PSR correspond to enhancements in the 
number of tagged photons available to the experiment. The agreement between the 
data and the simulation is excellent.

\subsection{Scaler rates and system deadtime}
\label{subsection:scaler_rates_and_system_deadtime}

\begin{figure}
\resizebox{1.00\textwidth}{!}{\includegraphics{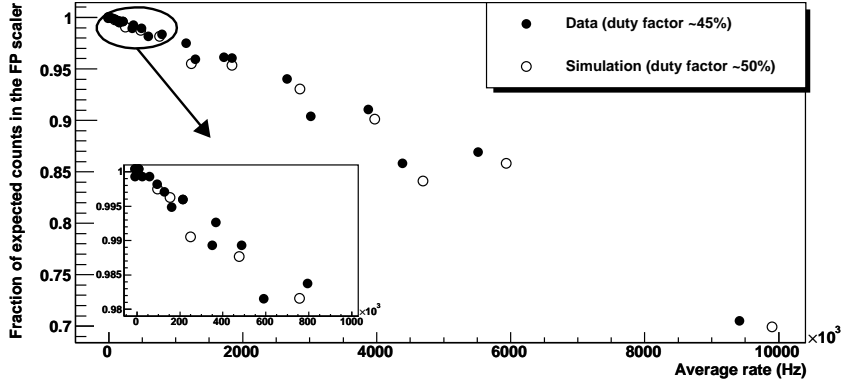}}
\caption{\label{figure:figure_14_recoil_electron_loss_effects}
Recoil-electron loss effects. A comparison between data (filled 
circles) and simulation (open symbols) of the fraction of post-bremsstrahlung 
electrons counted by the FP scalers as a function of 
average recoil-electron rate. 
The inset expands the region up to 1~MHz. See text for details.
}
\end{figure}

A rate-dependent study of the effects of deadtime in the FP electronics was 
performed using a standard 45\% duty-factor electron beam to determine the 
impact on the FP scalers. FP channels 5, 15, 25, and 35 were first tested 
individually, then OR-ed together in two groups of two (5 and 15, 25 and 35), 
and finally tested as a single group of four in order to conveniently increase 
the available rates. The counts in the individual channels were used to 
determine the actual numbers of counts reaching the FP scalers.  The fraction 
of counts reaching the FP scalers for the twofold and fourfold ORs was then 
monitored as a function of rate.
Figure~\ref{figure:figure_14_recoil_electron_loss_effects} shows data (solid 
circles) for the fraction of expected counts in the twofold and fourfold 
configurations as a function of average recoil-electron rate. Simulated values 
are shown for a duty factor of about 50\% 
(open circles). The inset shows the densely populated region from 0 to 1~MHz. 
Clearly, the simulation does an excellent job of predicting the loss of counts 
in the FP scalers due to deadtime effects.

\subsection{Prescaled focal-plane trigger}
\label{subsection:pre_scaled_focal_plane_trigger}

\begin{figure}
\resizebox{1.00\textwidth}{!}{\includegraphics{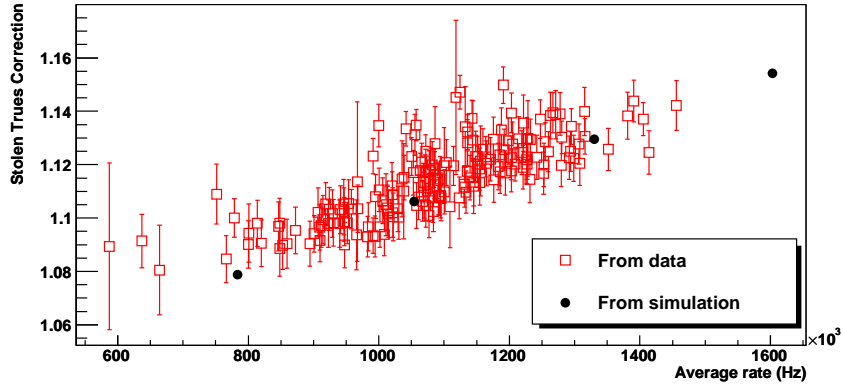}}
\caption{\label{figure:figure_15_pFP_comparison}
(Color online) A comparison between the stolen-coincidences corrections for 
the pFP trigger obtained from the data (open squares) and the simulation 
(filled circles).
}
\end{figure}

It is straightforward to compare the data collected using the pFP trigger to 
the simulation because the trigger source is one of the FP channels. The number
of stolen coincidences in the pFP trigger channel is the ratio of the number of
counts in the TDC prompt peak (recall Fig.~\ref{figure:figure_07_pFP}) to the 
total number of triggers. Since events with an output width of less than 11~ns 
in the coincidence overlap module do not register a stop in the FP TDCs, it is 
important to include events to the right of the prompt peak as stolen 
coincidences since these events were not recorded due to the limitations in 
the electronics setup. Using data, the stolen coincidences can be determined 
for the pFP trigger on a run-by-run basis (recall 
Fig.~\ref{figure:figure_07_pFP}). The stolen coincidences can also be 
calculated using the simulation at several recoil-electron rates. A comparison 
of the two approaches is presented in 
Fig.~\ref{figure:figure_15_pFP_comparison}.

\subsection{Stolen-coincidence correction}
\label{subsection:stolen_coincidence_correction}

\begin{figure}
\resizebox{1.00\textwidth}{!}{\includegraphics{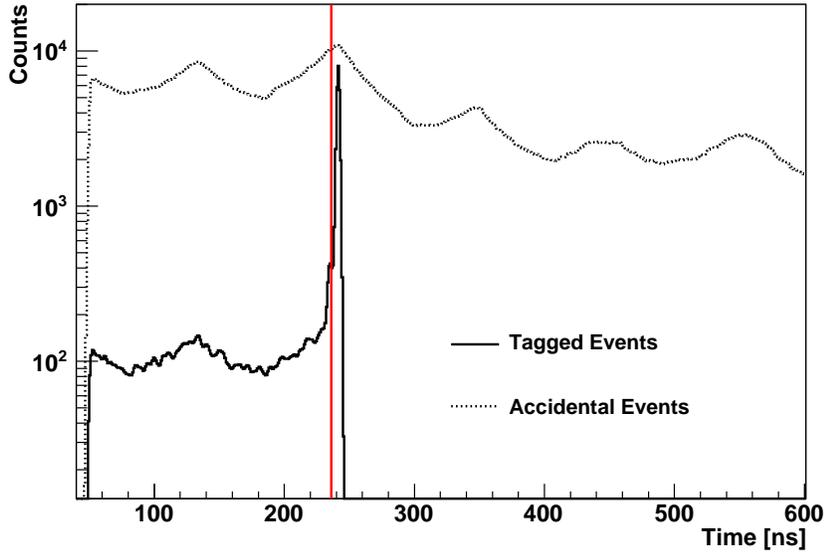}}
\caption{\label{figure:figure_16_stolen_methods}
(Color online) Spectra for FP channel 30 used to obtain the stolen-coincidence 
correction via Eq.~\ref{equation:schroder_correction}. The lower solid spectrum
corresponds to the tagged-photon TDC, while the upper dotted spectrum is 
simulated accidentals.  The prompt peak is located at $t_0$ = 240~ns. Stolen 
coincidences appear to the left of this peak. The red line indicates the 
location of $t_{\rm left}$. See text for details.
}
\end{figure}

A method for analytically calculating the stolen coincidences for a nearly 
continuous beam is given in Ref.~\cite{owens1990}. The stolen-coincidence 
correction is given by

\begin{equation}
\label{equation:stolen_coincidence_correction}
f_{\rm stolen} =
e^{R\tau},
\end{equation}
where $R$ is the electron rate and $\tau$ is the time between the start of the 
TDC and the arrival of the corresponding tagged electrons.  The 
stolen-coincidence correction can also be determined via 
the simulation. The simulated, tagged-photon TDC spectrum (see 
Fig.~\ref{figure:figure_16_stolen_methods}) shows both prompt events and 
stolen coincidences. The fraction of stolen coincidences determined using the simulated
tagged photons is given by
\begin{equation}
\label{equation:stolen_coincidence_correction2}
f_{\rm stolen}^{\rm sim,tag} =
\frac{N}{N_{\rm prompt}},
\end{equation}
where $N$ is the total number of events and $N_{\rm prompt}$ is the number 
of events arriving during the prompt window.

Several values of $R$ and $\tau$ were chosen as simulation inputs in order to 
compare the simulated stolen coincidences to those predicted by 
Eq.~\ref{equation:stolen_coincidence_correction}. The results are shown in 
Table~\ref{table:table_02_stolen_coincidence_correction}. Agreement is 
excellent.

\begin{table}
\caption{\label{table:table_02_stolen_coincidence_correction}
Comparison of the fraction of stolen coincidences obtained using 
Eq.~\ref{equation:stolen_coincidence_correction} and the simulation.}
\begin{center}
\begin{tabular}{cccc} \hline \hline
Rate$_{\rm inst}$ /MHz & $\tau$ /ns & $f_{\rm stolen}^{\rm calculated}$ & $f_{\rm stolen}^{\rm simulated}$ \\
\hline
     1.0 &         19 &                                 1.019 &                     1.018 $\pm$ 0.001 \\
     2.0 &         19 &                                 1.039 &                     1.036 $\pm$ 0.001 \\
     1.0 &        199 &                                 1.220 &                     1.211 $\pm$ 0.001 \\
     2.0 &        199 &                                 1.489 &                     1.477 $\pm$ 0.002 \\
\hline \hline
\end{tabular}
\end{center}
\end{table}

Another method \cite{hoorebeke1992} for determining the stolen-coincidence 
correction determines the correction from the accidental FP TDC 
spectrum \cite{hoorebeke1992} (see Fig.~\ref{figure:figure_16_stolen_methods}).
With this approach, the correction factor is given by

\begin{equation}
\label{equation:schroder_correction}
f_{\rm stolen}^{\rm accidentals} =
\frac{N}{N-N_{t<t_{\rm left}} },
\end{equation}
where $N$ is the total number of events and $N_{t<t_{\rm left}}$ is the number 
of events arriving prior to the coincidence peak (located at $t_0$). The 
stolen-coincidence correction was calculated for both a simulated accidental FP
TDC spectrum and the TDC spectrum of the tagged photons. The results are shown 
in Table~\ref{table:table_03_schroder_stolen_coincidence_correction}.
Agreement is excellent.

\begin{table}
\caption{\label{table:table_03_schroder_stolen_coincidence_correction}
Comparison of the of the stolen-coincidence correction obtained using 
Eq.~\ref{equation:schroder_correction} applied to the accidental TDC spectrum
and Eq.~\ref{equation:stolen_coincidence_correction2} applied to the tagged
photons TDC spectrum.}
\begin{center}
\begin{tabular}{ccc} \hline \hline
FP ch &  $f_{\rm stolen}^{\rm accidentals}$  & $f_{\rm stolen}^{\rm tagged}$ \\
\hline
      0 &         1.521 &                              1.519  \\
     10 &         1.567 &                              1.578  \\
     20 &         1.672 &                              1.670  \\
     30 &         1.739 &                              1.738  \\
     40 &         1.768 &                              1.785  \\
     50 &         1.900 &                              1.916  \\
     60 &         2.029 &                              2.056  \\
\hline \hline
\end{tabular}
\end{center}
\end{table}

\subsection{Ghosts correction}
\label{subsection:ghosts_correction}

The correction for the rate-dependent ghost effect differs from the 
stolen-coincidence correction in
that an exact analytical form for the correction does not exist. However, it is 
possible to compare the results of the simulation to the values used by 
Hornidge~{\it \mbox{et al.}}~\cite{hornidge1999,hornidge2000}. 
Unfortunately, it is impossible to replicate the exact parameters of the SAL 
configuration.  Additionally, the results given by Hornidge~{\it et al.} are 
run-averaged results. Still, a comparison of the ghost corrections is useful 
in evaluating the overall accuracy of our FP simulation.

In order to perform such a comparison, the beam conditions were reconstructed 
as carefully as possible to best reproduce the conditions of the SAL 
experiment. The beam rates (1.4 and 4.755~MHz) were chosen to match two of 
those listed by Hornidge~{\it et al.} The ghost correction is defined as

\begin{equation}
\label{equation:ghost_correction}
1/f_{\rm ghost} =
\sigma_M/\sigma,
\end{equation}
where $\sigma_{M}$ is the measured cross section and $\sigma$ is the real 
cross section. The correction factors predicted by the FP simulation are 
compared to the values obtained by Hornidge~{\it et al.} in 
Table~\ref{table:table_04_ghost_correction}.
The apparent disagreement in these results is likely due to undocumented 
differences in the two experimental setups (for example: discriminator modes,
pulse timing, TDC behavior, and beam structure). It is probably impossible to 
improve the agreement without completely recreating the precise conditions of the SAL experiment.

\begin{table}
\caption{\label{table:table_04_ghost_correction}
Comparison of the ghost correction obtained in Ref~\cite{hornidge1999} and our 
simulation.
}
\begin{center}
\begin{tabular}{ccc} \hline \hline
Rate$_{\rm inst}$ /MHz & $f_{\rm ghosts}^{\rm Hornidge}$ & $f_{\rm ghosts}^{\rm this~work}$ \\
\hline
    1.400 &                           1.019 &                            0.993 \\
    4.755 &                           1.076 &                            1.007 \\
\hline \hline
\end{tabular}
\end{center}
\end{table}

\subsection{Systematic Uncertainties}
\label{subsection:systematic_errors}

The systematic uncertainties in the ghosts and stolen-coincidence corrections 
was determined as a function of variations in (1) the FP discriminator pulse 
width, (2) the time profile of the stretched electron beam leaving the PSR, 
and (3) the electron rate in the individual FP channels.  

During the inaugural run periods at the TPF, the FP discriminator pulse widths 
were typically set to (50$\pm$1) ns\footnote{
Today, the FP discriminator pulse widths
are typically set to (25$\pm$1) ns.}.
In the simulation, the discriminator pulse widths were varied by $\pm$2 ns to 
see how the corrections were affected.  In this case, both the 
stolen-coincidence and ghosts correction varied by less than 0.5$\%$.

Understanding the effect of the time profile of the beam on the corrections was
more involved. Conceptually, the electron-beam profile after the PSR could be 
more uniform (``flatter" as a function of time representing a more continuous
beam) or less uniform (higher peaks and lower valleys as a function of time
representing a less continuous beam) than the profile used. To obtain a more 
uniform time profile, $[P(t)]^{0.75}$ was employed; the less uniform profile 
was given by $[P(t)]^{1.33}$. Exponents smaller than 0.75 or larger than 1.33 
resulted in simulated accidental TDC spectra that did not reproduce the data. 
The variation in the stolen-coincidence correction due to these extreme 
profiles was less than 2$\%$. The variation of the ghosts correction was less 
than 3$\%$.

The stolen-coincidence correction depends on the electron rate according
to Eq.~\ref{equation:stolen_coincidence_correction}. Further, a 
rate-independent method to extract the correction is presented in 
Eq.~\ref{equation:stolen_coincidence_correction2}.  The difference between 
these methods is typically 2\% or less. Using Eq.~\ref{equation:stolen_coincidence_correction}
and assuming the electron rate varies by $\sim$10\%, the correction is found to 
have a systematic uncertainty of $\le$ 5\%. Combining these results, the
variation of the stolen-coincidence correction as a function of electron rate
is taken to be 5$\%$.
The effect of the electron rate on the 
ghosts correction was investigated by varying the nominal electron beam rate 
during in the simulation by $\pm$25$\%$. The ghost 
correction varied by less than 1$\%$ over this range of beam rates.

A summary of the systematic uncertainties in the stolen-coincidence and 
ghosts corrections due to simulation parameter variations is presented in
Table~\ref{table:table_05_sys_uncertainties}.

\begin{table}
\caption{\label{table:table_05_sys_uncertainties}
Systematic uncertainties in the stolen-coincidence and ghosts corrections 
due to simulation parameter variations. 
}
\begin{center}
\begin{tabular}{ccc} \hline \hline
                                             & \multicolumn{2}{c}{effect on correction} \\
varied parameter                             & stolen coincidences &             ghosts \\
\hline
FP discriminator pulse width                 &               0.5\% &              0.5\% \\
time profile of the extracted electron beam  &                 2\% &                3\% \\
electron rate in FP channels                 &      5\% (see text) &                1\% \\
\hline \hline
\end{tabular}
\end{center}
\end{table}

\subsection{Carbon elastic scattering cross section for photons}
\label{subsection:carbon_cross_section}

\begin{figure}
\resizebox{1.00\textwidth}{!}{\includegraphics{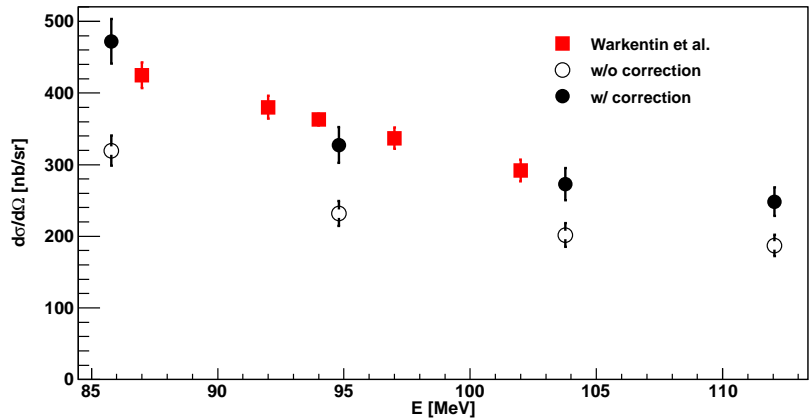}}
\caption{\label{figure:figure_17_carbon}
(Color online) Absolute differential cross-section data corrected for 
rate-dependent effects as outlined in this paper compared to published data. 
Statistical uncertainties only are shown. See text for details.
}
\end{figure}

As an illustration of the success of our Monte Carlo method for addressing 
rate-dependent effects, Fig.~\ref{figure:figure_17_carbon} presents a 
comparison between the absolute differential cross section recently obtained 
at the TPF at the MAX IV Laboratory and existing data published by 
Warkentin~{\it et al.}~\cite{warkentin2001} (filled squares) for elastic 
photon scattering from $^{12}$C at a lab angle of 60$^{\circ}$ performed at 
SAL. Error bars reflect statistical uncertainties only. The open circles show 
our cross-section data prior to correction for rate-dependent effects, while 
the filled circles show the same data after correction for rate-dependent 
effects.  The agreement between the two data sets is excellent. A summary of 
the rate-dependent corrections is presented in 
Table~\ref{table:table_06_all_corrections}.

\begin{table}
\caption{\label{table:table_06_all_corrections}
Stolen-coincidence and ghosts corrections applied to the $^{12}$C data 
shown in Fig.~\ref{figure:figure_17_carbon}. Non-negligible statistical 
uncertainties are listed.
}
\begin{center}
\begin{tabular}{ccccc} \hline \hline
photon energy / MeV               &           86 &           95 &          104 &          112 \\
\hline
Uncorrected Cross Section / nb/sr & 320 $\pm$ 21 & 232 $\pm$ 17 & 202 $\pm$ 16 & 187 $\pm$ 15 \\
Stolen Trues Correction           &         1.49 &         1.43 &         1.37 &         1.33 \\
Ghosts Correction                 &         0.99 &         0.99 &         0.99 &         1.00 \\
Corrected Cross Section / nb/sr   & 472 $\pm$ 31 & 327 $\pm$ 25 & 273 $\pm$ 22 & 248 $\pm$ 20 \\
\hline \hline
\end{tabular}
\end{center}
\end{table}

\section{Summary}
\label{section:summary}

Rate-dependent effects in the electronics used to instrument the tagger focal
plane at the Tagged-Photon Facility at the MAX IV Laboratory have been 
investigated using a dedicated Monte Carlo simulation. The Monte Carlo 
simulation incorporates the unique behaviors of each of the critical 
focal-plane instrumentation modules. Results have been compared to analytical 
calculations of these effects, as well as experimental data collected for a 
series of specialized tests of the rate-dependent response.  The simulation 
agrees very well with both. Further, the output of the simulation has been used
to normalize high-rate tagged-photon production data (1~MHz average and up to 
4~MHz instantaneous rates per focal-plane channel). Rate-corrected 
cross-section data are in excellent agreement with previous results obtained at
lower instantaneous rates.  We assert that this Monte Carlo simulation is of 
fundamental importance to the analysis of all experimental data from the 
Tagged-Photon Facility at the MAX IV Laboratory and, in principle, is adaptable
to any high-rate coincidence experiment.

\section*{Acknowledgements}
\label{acknowledgements}

This project was supported by the US National Science 
Foundation Grant No. 0855569, as well as The Swedish Research Council, the 
Crafoord Foundation, and the Royal Physiographic Society in Lund.
The authors gratefully acknowledge the Data Management and Software Centre, 
a Danish Contribution to the European Spallation Source ESS AB, for generously 
providing access to their computations cluster.

\bibliographystyle{elsarticle-num}

\end{document}